\documentclass[namedreferences]{SolarPhysics}
\usepackage{graphicx}

\newcommand{\be}{\begin{equation}}
\newcommand{\ee}{\end{equation}}
\newcommand{\beq}{\begin{eqnarray}}
\newcommand{\eeq}{\end{eqnarray}}

\begin{document}
\begin{article}
\begin{opening}

\title{EUV Analysis of a Quasi-Static Coronal Loop Structure}

\author{J.T. Scott$^{1}$, P. C. H. Martens$^{1,2}$, and D. McKenzie$^{1}$}

\runningauthor{SCOTT ET AL.}
\runningtitle{EUV Analysis of a Quasi-Static Coronal Loop Structure}

\institute{$^{1}$ Department of Physics, Montana State University,
                  Bozeman, MT, 59717, U.S.A.,
                  \email{scott@solar.physics.montana.edu}\\
           $^{2}$ Harvard-Smithsonian Center For Astrophysics, 60 Garden Street, Cambridge, MA 02138\\}

\date{Received April 7, 2011; accepted October 14, 2011 }

\begin{abstract}

Decaying active region 10942 is investigated from 4:00-16:00 UT  on February 24, 2007 using a suite of EUV observing instruments.  Results from {\it Hinode}/EIS, STEREO and TRACE show that although the active region has decayed and no sunspot is present, the physical mechanisms that produce distinguishable loop structures, spectral line broadening, and plasma flows still occur. A coronal loop that appears as a blue-shifted structure in Doppler maps is apparent in intensity images of  log(T) = 6.0-6.3 ions. The loop structure is found to be anti-correlated with spectral line broadening generally attributed to nonthermal velocities. This coronal loop structure is investigated physically (temperature, density, geometry) and temporally. Lightcurves created from imaging instruments show brightening and dimming of the loop structure on two different time scales; short pulses of 10-20 min and long duration dimming of 2-4 hours until its disappearance. The coronal loop structure, formed from relatively blue-shifted material that is anti-correlated with spectral line broadening, shows a density of $10^{10}$ to $10^{9.3}$ cm$^{-3}$ and is visible for longer than characteristic cooling times. The maximum nonthermal spectral line broadenings are found to be adjacent to the footpoint of the coronal loop structure. 

\end{abstract}
\keywords{Sun: Corona, Sun: EUV, Sun: Coronal Loops, Doppler velocity, Non-thermal spectral line broadening}

\end{opening}

\section{Introduction}\label{Introduction}

How and where the heating of the solar corona occurs remains a significant astrophysical question. A multitude of theories have been suggested for heating the corona. One way to probe the heating of this plasma is through observations of looplike structures that appear bright against the surrounding background and trace out magnetic structures. The closed magnetic field that brightens, deemed coronal loops, has been investigated with space-based instruments since the first X-ray observatory {\it Skylab}.

The early observations with soft-X-ray instruments showed long-lived hot ($> 2$ MK) loops. These loops seemed to agree with the static scaling laws developed by Rosner, Tucker, and Vaiana (1978). With the rise of new observing instruments, new views of coronal loops were obtained. The loops seen with the EUV instruments appear to have temperatures near 1 MK and densities greater than those predicted for static loops (Aschwanden, Schrijver, and Alexander, 2001). The apparent isothermality derived with filter ratios and constant loop widths led to the belief that elementary loop strands were resolvable with current instruments (Aschwanden and Nightingale, 2005; Aschwanden {\it et al.}, 2008; L\'opez Fuentes, D\'emoulin, and Klimchuk, 2008). Spectral analysis of coronal loops (with CDS) gave a conflicting result (Schmelz {\it et al.}, 2001). Loops were found to have a wide distribution in temperature and thus  considered multi-thermal. These observations  led to the question of whether the observed coronal loops are monolithic isothermal or multi-strand multi-thermal entities (Klimchuk, 2009). 

The approach in distinguishing between the hypotheses has been to estimate the physical parameters, and lifetime of an isolated loop structure and then compare the measurements to relevant coronal heating models. Previous studies (Aschwanden, Schrijver, and Alexander, 2001; Winebarger and Warren, 2005) show that these loops may not be in static equilibrium and display lifetimes longer than those predicted for radiative and conductive cooling. This was best explained by multiple strands heated sequentially (Warren, Winebarger, and Mariska, 2003; Ugarte-Urra, Winebarger, and Warren, 2006). A recent attempt at reconciling both the multithermal/isothemal debate has been put forth by Klimchuk, (2009). This model uses the nanoflare idea proposed by Parker. Klimchuk, (2009) suggests that the thermal distribution of the isolated loop structures is related to the duration of the nano-flare events. More accurate estimates of the thermal distribution of these structures are now obtainable using the state-of-the-art {\it EUV Imaging Spectrometer} (EIS), (Culhane {\it et al.}, 2007). The thermal distribution of an isolated looplike structure is one of five key parameters important for constraining this and other loop heating models. Density, lifetime, flows, and intensity profile complete the outlined parameters (Klimchuk, 2009). Few observations of all these parameters have been accomplished/published for an isolated loop structure.  Warren {\it et al.} (2010) measures these parameters from an apparently cooling loop structure. They observe a structure first in the XRT X-ray images then in TRACE 195 \AA\ and later 171 \AA\, indicating cooling.

It is thought that loops observed in EUV are generally seen only in their cooling phase (Winebarger, Warren, and Seaton, 2003). Investigations show that large, cool extended structures lay above hot dense cores of active regions (Mason {\it et al.}, 1999). The cooler structures ($< 2$ MK) tend to brighten more transiently while the hotter cores appear to be steady. Understanding how these hot, cool, and warm structures fit into the larger active region is important for understanding how the active region is heated. Recent studies (Del Zanna, 2008; Ugarte-Urra, Warren, and Brooks, 2009) investigate the relationship between these loop structures and the active regions they are embedded in. Ugarte-Urra, Warren, and Brooks (2009) observe core active region loops that cool through the entire temperature range of the data set (2.5 - 0.4 MK). They also find cool loops on the periphery of the active region that are not associated with the cooling of multi-million degree plasma. Del Zanna (2008), using EIS observations, showed line-of-sight relative red shifts in cool material with blueshifted hotter material in regions with sharp boundaries and relatively low-density regions. While blueshifted material is generally associated with low-density and low-intensity regions, we find a blueshifted dynamic coronal structure that displays both higher density and higher intensity than the surrounding background. Doschek {\it et al.} (2007) investigated nonthermal widths of Fe line profiles within active regions. Their study showed that the largest widths were not correlated with intensity, but found that the larger widths could be concentrated in blueshifted  outflow regions. Further analysis by Doschek {\it et al.} (2008) showed that the outflow regions of active regions are generally correlated with an increased line width.  Antolin {\it et al.} (2010) distinguish between Alfv\'en wave heating and nanoflare reconnection heating by characterization of the flow velocities of coronal loops. They find redshifts for Alfv\'en wave heating and  blueshifts for nanoflare reconnection heating.

Recent studies of coronal active regions find observed properties of the regions in general or over large areas of the active regions. We use multiple solar observing instruments to observe and isolate a coronal loop structure. Many coronal loop studies are focused on one or a few of  properties of many loop structures while concentrating on a specific heating model. We obtain a number of properties relevant to multiple coronal heating models for an isolated structure. Our investigation expands on previous studies of active region phenomena in general and looks at a smaller region within the active region, at the main component of the corona, the coronal loop.

 This study investigates the physical properties, flows, line broadening, and lifetime of an isolated coronal structure in the core of an active region. The surrounding active region is examined. In Section 2 we outline the instruments and observations of the active region. We discuss the spectral analysis with EIS and investigation of the complementary images provided by STEREO and TRACE in Section 3. The results of this analysis are given in Section 4. We report the basic parameters of the loop.  It is shown in Section 4 that warm loop structures in the core of an active region can show relative blueshifts along their lengths with an absence of spectral line broadening.

\section{Observations}

\begin{figure}
\begin{center}
\includegraphics[totalheight=3in] {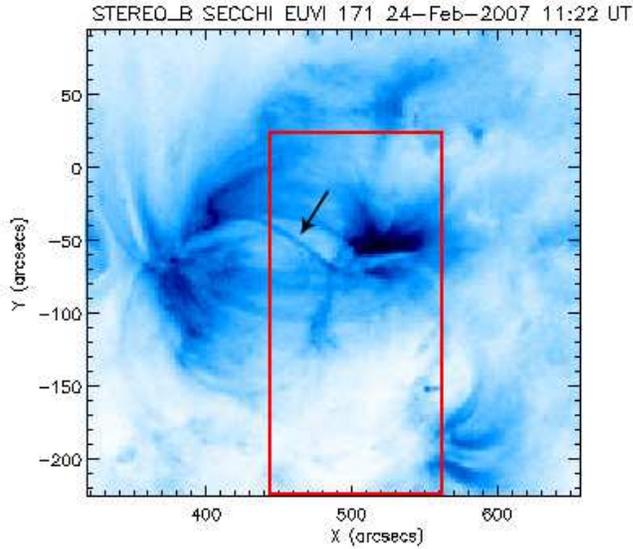}
\caption{ STEREO 171 \AA\ image of AR 10942 with EIS FOV delineated by the box. The arrow indicates the loop isolated for investigation. }\label{fig:stereo_w_eis}
\end{center}
\end{figure}

Spectra from the EIS spectrometer on {\it Hinode} and EUV images from STEREO and  TRACE are utilized in this study. Below, we present an outline of the observations of structures in AR 10942 from about 4:00 -16:00 UT February 24, 2007. No significant flare events occurred during this time, so the loop structure investigated in this study could be considered a quiescent loop. Figure \ref{fig:stereo_w_eis} shows the active region and loop structure, with a box indicating the EIS field of view.

The EIS instrument onboard {\it Hinode} (Culhane {\it et al.}, 2007) observed AR $10942$ with a $1$ arcsec slit scanning mode with $1$ arcsec steps. EIS began rastering the region with 20 seconds exposures at 11:21 UT and completed the raster at  12:03 UT. The raster imaged a $120$ arcsec $\times\ 120$ arcsec area close to the solar equator. Spectra from the two wavelength ranges (171 -212 \AA\ and 245 -291 \AA) allow for the observation of many ion lines spanning a large temperature range. Data from $17$ spectral windows were collected. The standard solarsoft routine eis\_prep was used to prep and calibrate the raw data. The instrumental units (counts) were converted to physical units. The spectral windows were then fit with Gaussians to determine the total intensities, peak wavelengths and widths of the ion line profiles. Careful attention was paid to identify and to account for blends within these windows. The EIS images where then corrected for: offsets between the high and low wavelength ranges, orbital variation, and spectral rotation (see Young {\it et al.}, 2009).

The {\it Transition Region and Coronal Explorer} (TRACE) instrument (Handy {\it et al.} 1999) is capable of imaging a $512$ $\times\ 512$ arcsec field of view with CCD pixels that correspond to $0.5$ $\times\ 0.5$ arcsec. Observations of AR $10942$ mainly consist of $256$ arcsec $\times\ 256$ arcsec  images with fewer $512$ arcsec $\times\ 512$ arcsec,$1024$ arcsec $\times\ 1024$ arcsec and $118$ arcsec $\times\ 256$ arcsec images. Unfortunately the $171$ \AA\  waveband was the only EUV filter imaging during the considered time.  The TRACE data were prepped and calibrated using trace\_prep. The images were normalized by the exposure time and inter-co-aligned. From about 8:00-16:00 UT the TRACE instrument took 228 images in the $171$ \AA\ waveband with a cadence of roughly 30 seconds with a few gaps of $\approx$ 10-20 minutes.


The {\it Sun-Earth-Connection Coronal and Heliospheric Investigation} (SECCHI) extreme ultraviolet (EUV) imager (EUVI) instruments onboard the {\it Solar Terrestrial Relations Observatory} (STEREO) are normal incidence telescopes like TRACE. The EUVI (Howard {\it et al.}, 2008) observes the full Sun with a pixel size of about 1 arcsec. There are two STEREO spacecraft, both with EUVI instruments that image in four wavebands ($171$, $194$, $284$, and $304$ \AA). Data from the STEREO {\it Ahead} and {\it Behind} EUVI 171, 195, 284 \AA\ wavebands were utilized for this study. The solarsoft routine secchi\_prep, included in the STEREO package, was used to prep and normalize the data. The EUVI 171 and 194 \AA\ images were taken with a cadence of $\approx$ 10 min collected from 4:06-16:06 UT, the $284$ \AA\ images had a cadence of $\approx$ 45 min and so they were not included in the light curves.

\section{Analysis}

The physical parameters of the coronal loop structure are first derived using spectral data from EIS. Density, temperature, and emission measure are found from background subtracted ion emission intensities. The loop structure is isolated and then the loop geometry is interpolated to match the pixel geometry for all the data. A loop structure may cross a group of pixels on their diagonals or at an angle to the pixel orientation. The image data is then interpolated such that the isolated loop lies parallel the pixel orientation and not at a diagonal. This procedure straightens the curved structure and aligns the pixel rows to be perpendicular to the long axis of the structure.

\subsection{EIS} 

The data from 17 spectral windows were recorded. The spectral range of the windows was investigated for corrupted data, bad spectra or the lack of a resolvable emission line. These pixels were omited from the study. The ion emission lines selected for this study are shown in Table I. 

\begin{center}
\begin{table} 
\caption{Spectral Lines Observed with EIS.  $\lambda_{pred}$ is the predicted wavelength for each transition. $T_{pf}$ is the peak formation temperature of the ion. The Ca \small{XVII} line is blended with O {\small V} (log(T) = 5.4 and Fe {\small XI} (log(T) = 6.2) and was solely used for visual comparison.}

\begin{tabular}{|l|l|l|}
\hline\hline
$ION$& $\lambda_{pred} ($\AA$) $ & $T_{pf}(log_{10}) K $\\
\hline
He \small {II} & $256.320$ & $4.90$ \\
Fe {\small VIII} &  $186.600$ & $5.60$  \\
Si {\small VII} &  $275.400$ & $5.80$  \\
Ca {\small XVII} &  $192.820$ & $6.7$ \\
Fe {\small XII} &  $186.854+186.887$ & $6.10$  \\
Fe {\small XII} &  $195.119$ & $6.10$  \\
Fe {\small XIII} &  $202.040$ & $6.20$  \\
Fe {\small XIII} &  $203.830$ & $6.20$   \\
Fe {\small XIV} &  $264.200$ & $6.30$  \\
Fe {\small XIV} &  $274.200$ & $6.30$  \\
Fe {\small XV} &  $284.160$ & $6.40$ \\
Fe {\small XVI} &  $262.980$ & $6.40$ \\
\hline

\end{tabular}
\end{table}\label{tab1}
\end{center}

\begin{figure}[h]
\begin{center}
\includegraphics[totalheight=4in]  {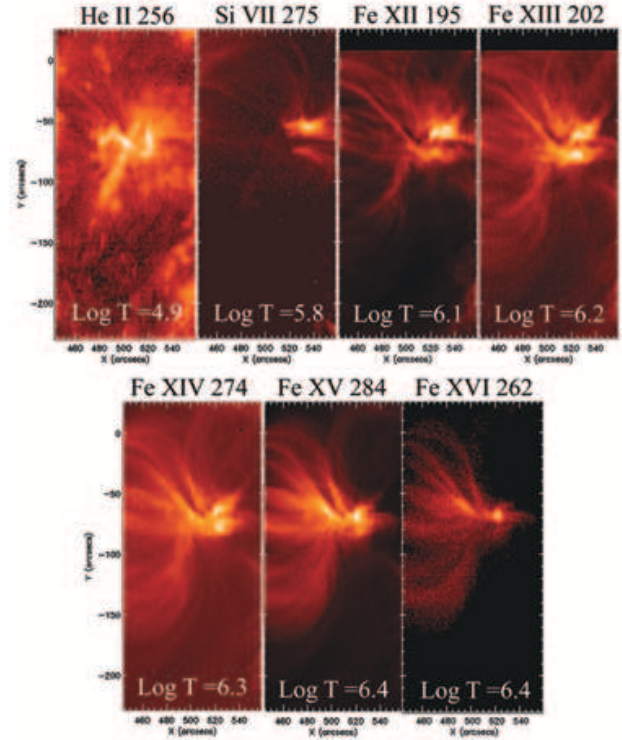}
\caption{EIS intensity images of decaying AR 10942. Temperatures  in Log(T) indicate the emission line’s formation temperature. Different features dominate different temperature ranges.  The investigated loop appears in Fe {\small  XII} and Fe {\small XIII}. As the temperature increases more diffuse interconnecting loops appear. }\label{fig:EIS_int}
\end{center}
\end{figure}

The spectral data for the emission lines were fit with Gaussians in order to determine total intensity and peak wavelength of emission for all selected lines. Figure \ref{fig:EIS_int} shows intensity images of the fitted emission lines. The Fe {\small XII} 195.12 profile is blended with Fe {\small XII} 195.18. Similarly, Fe {\small XIII} 203.82 and 203.797 and Fe {\small XII} 203.728 are blended in a spectral window. Double and triple Gaussians fits with a linear background were performed on these profiles to account for blending. A fitting routine was created to perform fits to the blended spectra utilizing constraints on the parameters of the fits for selected ions. The self blended Fe {\small XIII} 203 spectral window was fit using three Gaussians with equal widths, with the  value allowed to vary. The Fe {\small XII} 203.797 centroids were fixed relative to the Fe {\small XIII} 203.828 \AA\ line. A ratio of 0.40 for the Fe {\small XIII} 203.797/203.828 was used to constrain the peak values. This is reasonable since the density is found to be greater than $log N_e\ > 9.0$.  Similarly, the blended Fe {\small XII} 195 window was treated with two Gaussians. The 195.12 and 195.18 line widths are constrained to be equal and the 195.18 centroid forced to 0.06 \AA\ on the long wavelength side of the 195.12 centroid. A detailed examination of the relevant atomic data and density measurement techniques can be found in Young {\it et al.} (2009). The fitted spectral data from density sensitive lines were then corrected for spatial offset due to ``grating tilt'' (appendix C, Young {\it et al.} 2009).  Line-of-sight velocity maps were created from the fitted peak wavelengths. The FWHM ( \AA\ ) of the fitted line profiles were used to produce maps of the spectral line broadening.

 The loop structure was isolated and the loop geometry was interpolated to match the pixel geometry; effectively straightening the loop structure and aligning the pixels parallel to the long axis of the structure. Crosscuts of the straightened loop structure were then fitted with Gaussian profiles to determine background subtracted intensities along the isolated structure. The background subtracted from these profiles is linear in space. The fits to the cross axis profiles give background subtracted intensity and it is assumed that the FWHM of the Gaussian fit is the ``diameter'' of the structure. The intensity along the structure’s axis is then determined from the cross-axis intensity profile fits. Density is determined from ratios of the density sensitive emission lines, Fe {\small XII} 195.12, 195.18, 186, and   Fe {\small XIII} 202, 203. The emissivities for the wavelengths of each ion were used to create a density sensitive curve based on the background subtracted intensity ratios. These methods utilize version 6.0 of the CHIANTI atomic database (Dere {\it et al.}, 2009).

\subsection{TRACE}
The TRACE 171 \AA\ data were inter-co-aligned and co-aligned with the EIS images (Figure \ref{fig:EIS_int}). The Fe {\small X} 184 (log(T[K]) = 6.0)  and Fe {\small XII} 195 (log(T[K]) = 6.1) EIS images were used for the co-alignment. The loop structure length was determined from these images. Surface curvature between east/west footpoints and circular geometry were considered in the calculation of length.  After co-alignment the TRACE images were interpolated and straightened with the same method as used with the EIS data. Cross-axis intensity profiles of a section of the straightened loop were then analyzed to determine a linear background and actual intensity. The same method used with the EIS data, fitting Gaussians to the profiles, was used. A background subtracted intensity was determined for each of the images along with the average intensity within a small box that included a section of the structure. These intensities, background subtracted and averaged, were recorded along with their times of exposure to produce light curves.

\subsection{STEREO EUVI}
\begin{figure}[h]
\begin{center}
\includegraphics[totalheight=4in]  {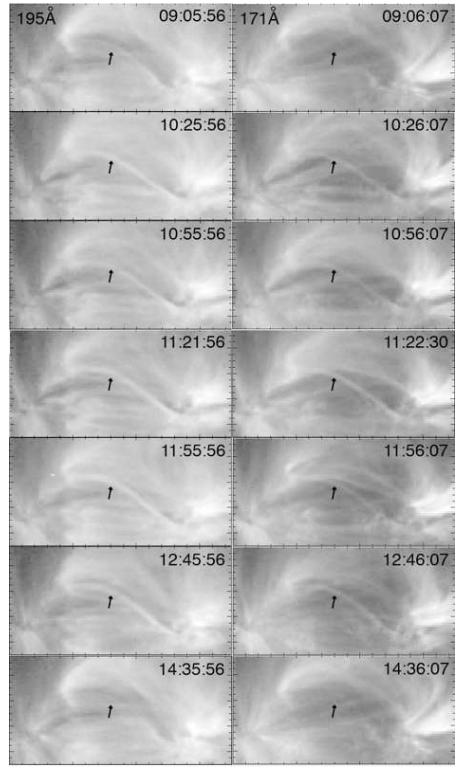}
\caption{STEREO 195 and 171 \AA\ images of AR 10942 for various times. The investigated loop structure is indicated by the arrows.}\label{fig:snapshots}
\end{center}
\end{figure}

The STEREO Behind 171 and 195 \AA\ images were first inter-co-aligned, then co-aligned with the EIS and TRACE images. The images were then interpolated in the same way the EIS and TRACE images were. Figure \ref{fig:snapshots} shows the core region of the active region for various times during the lifetime of the loop structure.  Cross-axis profiles of a co-spatial section of the structure were analyzed for background subtracted intensities and structure width. Light curves of 171 and 195 \AA\ were created from the background subtracted intensities taken from images during $\approx$ 04:00-16:00 UT. 

Images from both the STEREO Ahead and Behind 171 \AA\ were co-aligned and then used for 3D spectroscopic reconstruction (see Aschwanden {\it et al.}, 2008 for details) of the loop structure.

\section{Results}

\begin{figure}[h]
\begin{center}
\includegraphics[totalheight=4in]  {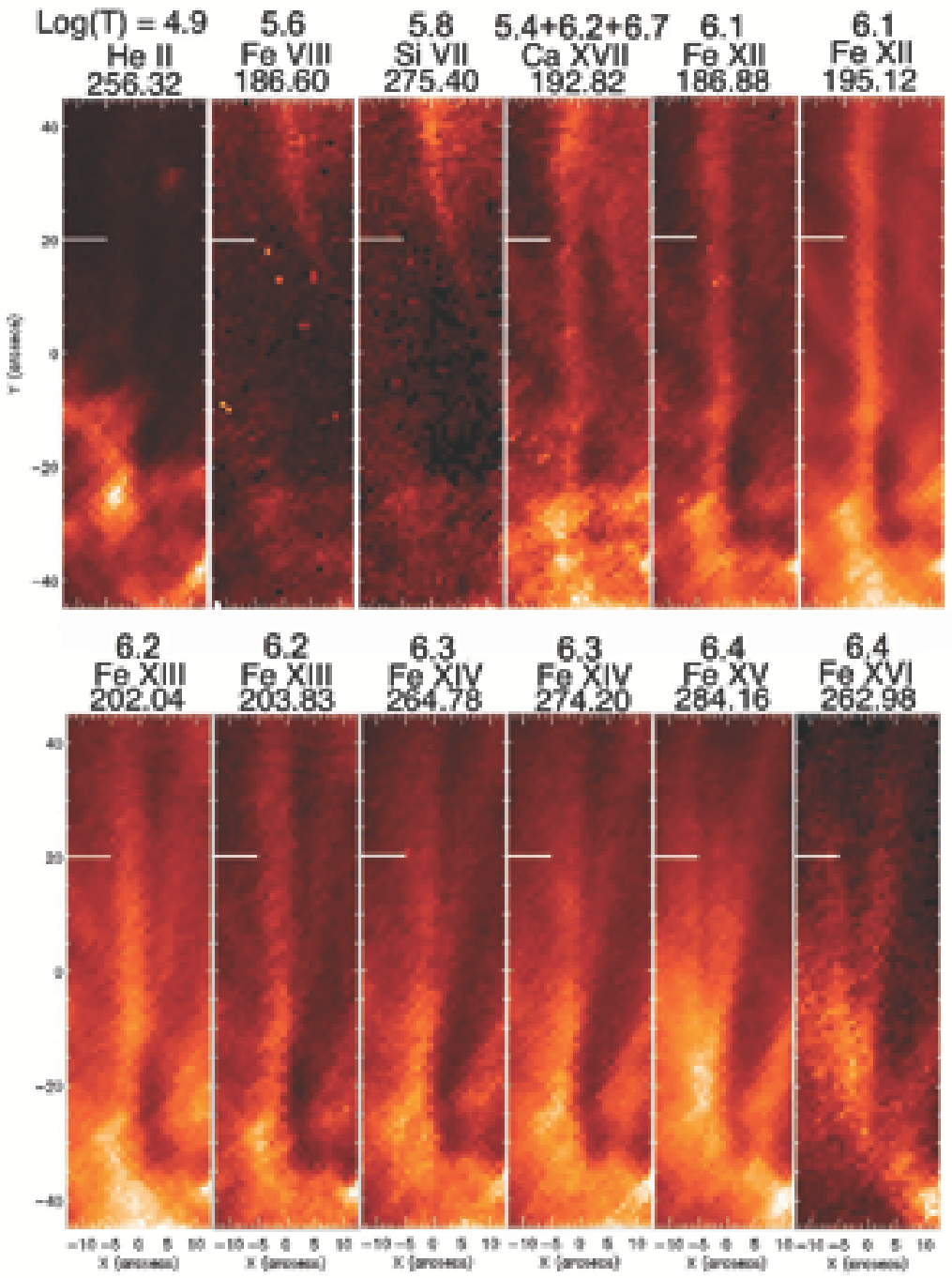}
\caption{EIS intensity images of interpolated loop. The loop structure, pointed out by the white bar, only appears in images formed at log(T) = 6.0-6.3. Structures that appear to share the same footpoint are evident in other temperature images, but the extended features are not co-spatial. The Ca {\small XVII} line is blended with O {\small V} (log(T) = 5.4) and Fe {\small XI} (log(T) = 6.2) and was solely used for visual comparison.}\label{fig:EIS_loops}
\end{center}
\end{figure}

The EIS spectroscopic observations of this 
active region show different dominant structures at different temperatures (Figure \ref{fig:EIS_int}).  The low temperature ions Fe {\small VIII} and Si {\small VII} both show none of the investigated structure and the footpoint region is not distinguishable from the images (Figures \ref{fig:EIS_int}, \ref{fig:EIS_loops}). They do show the apparent legs of loops that have footpoints separated from the loop structure analyzed. However the He {\small II} (log(T) = 4.9) map shows a bright region co-spatial with the footpoint of the investigated structure (Figures \ref{fig:EIS_int}, \ref{fig:EIS_loops}). The ions with formation temperatures log(T) = 6.0-6.3 show the loop that was investigated. The footpoint region appears bright and the loop could be distinguished from the background. We emphasize that the maximum intensity of the structures was found to be only 20-36 percent of the total signal in all of the imaging instruments and 35-50 percent of the signal from the spectrometer. The legs of two sets of loops can also be identified (brightest features in Si {\small VII} Figure \ref{fig:EIS_int}) and seem to be co-spatial with those seen in the temperature images from log(T) = 5.8-6.3. The higher temperature images of Fe {\small XV} and Fe {\small XVI} also show these legs, but less significantly in the Fe {\small XVI} where the footpoint is less obvious and the dominant emission is located farther up the leg of the structures. These temperature images do not clearly show the analyzed loop, but show bright structures similar in length and brightness to the structures in the log(T) = 6.0-6.3 temperature ranges. These structures appear to share the same footpoint region.

The interpolated image in Figure \ref{fig:EIS_loops}, after background subtraction, reveals a loop structure with a width close to the resolution limit of the observing instruments. STEREO and TRACE  images show  the structure did not change its apparent width during the time it was observed (about five hours). There is a slight decrease from the average in width for TRACE near the moments of appearance and disappearance. The loop widths calculated from Gaussian fits to the cross-loop profiles for the EIS and STEREO observations were on average $1.5 \pm 0.5$ arcsec while the TRACE apparent widths show $1.0 \pm 0.5$ arcseconds. This is easily explained by the difference in the resolution of the instruments. The TRACE pixels are 0.5 arcsecond and the EIS and STEREO pixels are around $1$ arcsecond. For both cases the loop structure's apparent width is at or below the resolution of the instrument. These widths can only then be viewed as an upper limit, since the structure does not appear to be resolved and is viewed at the limits of resolution by all instruments. Discussion of this can be found in DeForest (2007). The TRACE cross-loop profiles infrequently revealed up to three smaller maximums on top of the broader Gaussian profile.  This was found with the same infrequency in the STEREO profiles, but with two smaller maxima. This suggests multiple strands, below resolution, appearing as one structure.

Light-curves show flux variations for the 171 and 195 wavebands during the lifetime of the structure. While there are structures that brighten and fade completely with similar topology, the structure investigated remained distinguishable by our methods for just under 5 hours. Before the maintained appearance of the investigated feature, both wavebands display a structure in the same location as the investigated loop structure.  Intermittent brightenings occur for loop structures very close to the topology of the isolated loop structure as well. The flux variations of the investigated structure appear on the 10-20 minute scale. Shimizu (1995) investigated X-ray transient brightenings and found durations of 2-7 minutes. More recently Ugarte-Urra , Warren, and Brooks (2009) investigated loop lifetimes for temperatures 0.4 − 2.5 MK and higher. They found transient brightenings and loop lifetimes around 10-20 minutes in duration. Loops that appear to recur with the same topology were noted. They suggest that there are two main loop populations: core loops that appear at several million K and then cool to transition region temperatures and the second population consisting of cool peripheral loops that only reach temperatures of 1.3 MK. In both cases, they find time scales of tens of minutes, which is in agreement with our results. 

The EIS instrument rasters solar  west-east (right to left in Figure \ref{fig:EIS_int}) in the $x$ direction and considering 20 second exposures the region that contains the structure is exposed for 24 minutes (11:40-12:04). It is reasonable to assume that the loop may cool  during this time and appear in the lower temperature images later. The time of exposure increases in the $y$ direction for Figure \ref{fig:EIS_loops} and in the negative $x$ direction for Figure \ref{fig:EIS_int}. The apparent footpoint in the log(T) = 6.1-6.3 is shared by other structures in the log(T) = 6.4 images so it is unclear if the footpoint of the investigated structure is image by the higher temperature wavebands. Near the top  ($y=40$) of the log(T) = 5.6-5.8 images in Figure \ref{fig:EIS_loops}, a bright area coincides with the apex of investigated structure, which can be interpreted as the loop structure cooling into the lower temperature wavebands, but close inspection revealed that the apparent structure in the lower temperature images is separated topologically and the two structures intersect in the line of sight. The loop structure appears first in STEREO 195 \AA\ and shows an increase in flux during the EIS exposure times,  which is congruent with the structure appearing in the Fe {\small XII} and Fe {\small XIII} bands for the time EIS was rastering the loop. Figure \ref{fig:snapshots} and the light-curves show the structure is maintained in 195 \AA\ for this time period.

The loop structure is distinguishable in images from both STEREO Ahead and Behind, allowing for stereoscopic reconstruction. The stereoscopic reconstruction confirmed the estimated max loop half-length of 69 Mm,  and indicates that the legs of the loop are highly inclined towards the core of the active region.

\subsection{Density}
\begin{figure}[h]
\begin{center}
\includegraphics[totalheight=3in]  {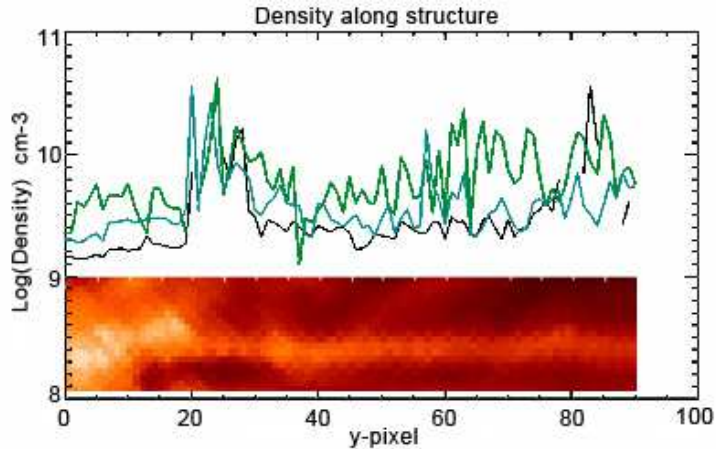}
\caption{Log Density along the loop structure. Turquoise is Fe {\small XII} 186/195.12, Black is Fe {\small XIII} 202/203, Green 195.12/195.18 Background subtraction begins at $y = 20$.}\label{fig:Density_Along}
\end{center}
\end{figure}

A density map, made using the fitted intensities, reveals the loop structure as an increase in density from the background. The background subtracted intensity values along the loop for the Fe {\small XII} and {\small XIII} ions show a decrease in density from $10^{10}$ to $10^{9.3}$ cm$^{-3}$ along the length of the feature. Figure \ref{fig:Density_Along} shows the density decrease according to all density sensitive ratios. However the Fe {\small XII} densities are consistently higher than those derived with the Fe {\small XIII} ratio. This is a known discrepancy (Young {\it et al.}, 2009). 

Emission measures for the densities found using EIS were calculated, using the observed path length of 1 Mm. The emission measures were compared with those derived from the TRACE 171 \AA\ fluxes. Filling factors of 0.047 near the footpoint to 0.09 near the apex were found from these comparisons.

\subsection{Temperatures from EM Loci plots} 

The determination of density along the structure allows for an estimate of the average temperature along the line of sight for locations along the structure. Co-spatial intensity values from emission lines, after background subtraction, are used to derive Emission Measure (EM) loci plots. Densities from the Fe {\small XII} 186/195.12 density sensitive ratio were used. For each emission line, the observed intensity is divided by the contribution functions calculated with CHIANTI. The functions for each line are plotted together. The average temperature is then determined from the highest concentration of intersections. Isothermal plasma along the line of sight would yield an EM-loci plot were all the curves intersected at the same location.

Some emission lines were not significant after background subtraction was applied to the loop structure (Si {\small VII}, Fe {\small XVI}, and Fe {\small XV}). A minimal value related to the background was used in order to include these lines in the temperature analysis. These high and low temperature emission lines constrain the EM-loci plot. Mostly Fe ions were considered, to avoid abundance uncertainties.
\begin{figure}
\begin{center}
\includegraphics[totalheight=2in] {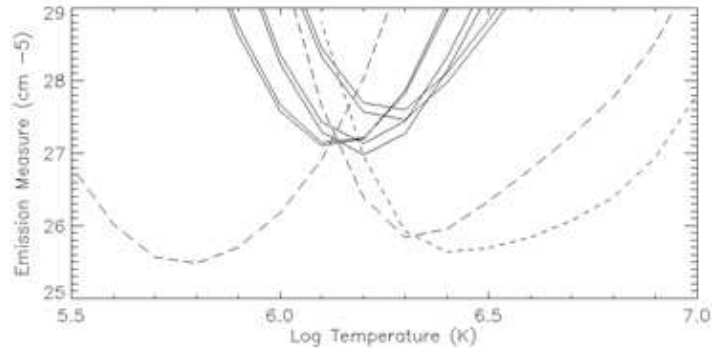}
\caption{EM Loci plot of pixel 24 along the loop. The dashed curves show the high and low temperature lines that were not significant after background subtraction. They are shown to impose constraints at high and low temperature ends. The median temperature for this pixel is log(T) = 6.27.} \label{fig:em_loci}
\end{center}
\end{figure}

\begin{figure}
\begin{center}
\includegraphics[totalheight=2in] {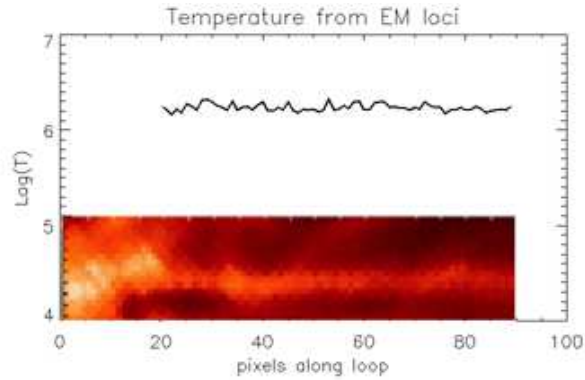}
\caption{A plot of temperature along the loop derived from the EM loci plots.} \label{fig:em_loci_temp}
\end{center}
\end{figure}

An estimate of the range of plasma temperatures along the line of sight was made from the EM-loci plots. The EM loci plots do not indicate an isothermal plasma, however they show a consistent maximum emission value at a temperature of log(T) $\approx$ 6.2 and a temperature width of log($\sigma$) = 5.56. The temperature width was found from the full width at half max of a Gaussian profile fitted within the EM curves. The EM loci plot from pixel 24 along the isolated structure is shown in Figure \ref{fig:em_loci}. 
A median temperature was derived from the intersections of the emission measure curves. This is shown in Figure \ref{fig:em_loci_temp}.  The plots differ very little along the loop structure and seem to indicate isothermality along the axis of the loop.  However, one should be cautious when interpreting these EM Loci plots since after background subtraction only density sensitive lines remained, delineating the structure (Fe {\small XII}, Fe {\small XIII}, and Fe {\small XIV}). These lines are non-ideal for EM analysis, but they were the only left to perform the EM loci analysis.

\subsection{Lightcurves}

\begin{figure}[h]
\begin{center}
\includegraphics[totalheight=4in]  {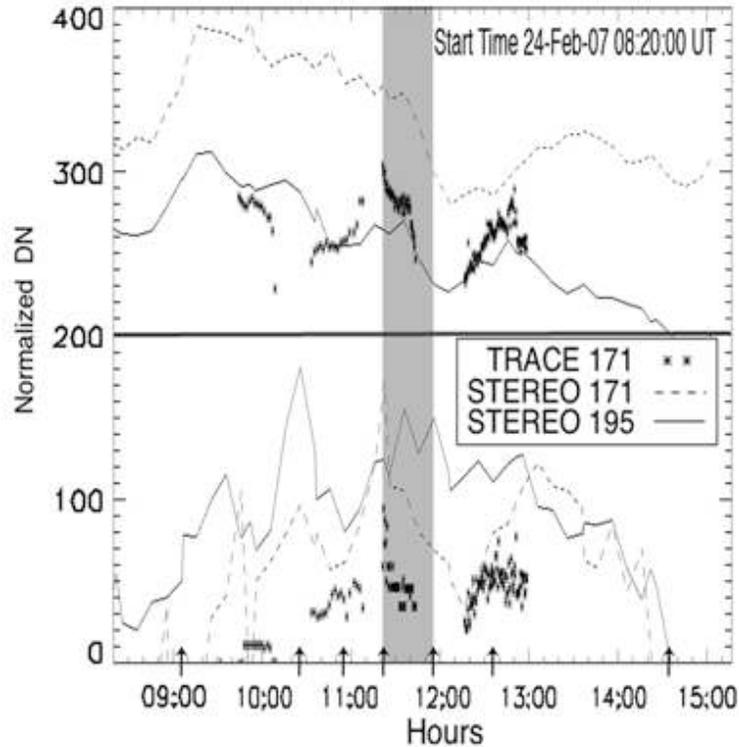}
\caption{Lightcurves from STEREO 171 (dashed), 195 (solid), and TRACE 171 (asterisk)  with EIS rastering time shaded. Above horizontal line: average intensities. Below horizontal line: background subtracted intensities. The plot begins at 08:20:00UT. Arrows indicate the times of the images from Figure 3.}\label{fig:Lightcurves}
\end{center}
\end{figure}

Co-aligned images from higher resolution filter imagers TRACE and STEREO show the same structures as their spectral counterparts. The background subtracted intensity values from cross loop profiles for a 10 hour period of 171 and 195 \AA\ data show an interesting lightcurve (Figure \ref{fig:Lightcurves}  shows the last 5 hours).
 The structure brightens and fades in all wavebands, but with two different time scales. The loop is just detectable in both channels at 9:00 UT then around 10:20 UT both 171 and 195 \AA\ show an increase in intensity. The 195 \AA\ passband experiences its maximum with a duration of 10 min. The 171 \AA\ channel shows a similar duration of 10 min. Then the structure dims simultaneously around 10:25 UT in both channels and begins to brighten again just before 11:00 UT. The 171 \AA\ lightcurve peaks for a second time at 11:21 UT. This is the largest peak recorded by the 171 \AA\ channel. This peak appears for a slightly shorter duration than the previous 195 \AA\ peak. Meanwhile the 195 \AA\ intensity continues to brighten steadily coming to a second, lesser, maximum at 11:35 UT (10 mins after the 171 maximum) and slowly, steadily, fading out until 14:35, where the structure cannot be distinguished anymore. After the 171 \AA\ intensity maximum the 171 \AA\ intensities show similar behavior as the 195 \AA\ lightcurve, but delayed by an hour. The loop fades then slowly steadily brightens to a second, lesser maximum and slowly fades out just before the 195 \AA\ intensity. The curves above the horizontal line at 200 (normalized DN) show the average intensity within the box selected for cross-loop investigation. These curves show the importance of isolating the structure from the background for every profile in the timeseries. The shaded area of the plot indicates the rastering duration of the EIS observations.

\subsection{Velocity}

\begin{figure}[h]
\begin{center}
\includegraphics[totalheight=4in]  {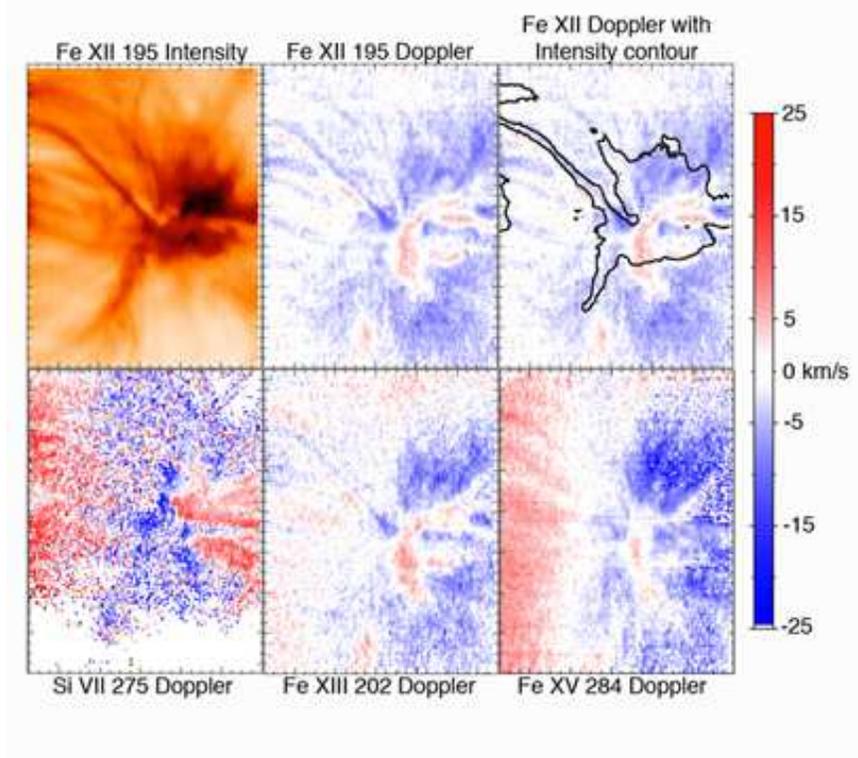}
\caption{EIS images of the 195 \AA\ intensity and relative Doppler velocities for Si {\small VII}, Fe {\small XII}, {\small XIII}, and {\small XV}. Upper right panel shows a Fe {\small XII} 195 \AA\ velocity image with intensity contours overlaid. The structure cannot be seen in the Si {\small VII} or Fe {\small XV} Doppler images.}\label{fig:Doppler}
\end{center}
\end{figure}

Doppler maps of the region in temperatures where the analyzed structure is seen reveal line-of-sight relative blue shifts along the entire structure. A Doppler map from Fe {\small XII} 195 with the intensity contours overplotted is shown in Figure \ref{fig:Doppler}. A small region, selected using the intensity images to encompass the footpoint region, show a blueshifted velocity ( -15$\pm\ $2 km s$^{-1}$) profile relative to the entire image and relative to a quiet region near the bottom of the image. The Fe {\small XIII} 202 \AA\ Doppler velocity for the region was ( -8$\pm\ $2 km s$^{-1}$).  This blueshifted velocity corresponds to a mass flow of around 4.2 $\times 10^{7}$ g s$^{-1}$ through the observed loop crosssection. The legs or two sets of footpoints  that appear in a large range of temperatures have LOS relative redshifts, but are surrounded by relative blueshifts. This is most notably seen in the Si {\small VII} velocity image (Figure \ref{fig:Doppler} lower left panel). The blueshifts from high temperature (T $> 1.25$) ions are present in the outer regions of the AR.

\subsection{Spectral Line Broadening}

\begin{figure}[h]
\begin{center}
\includegraphics[totalheight=4in]  {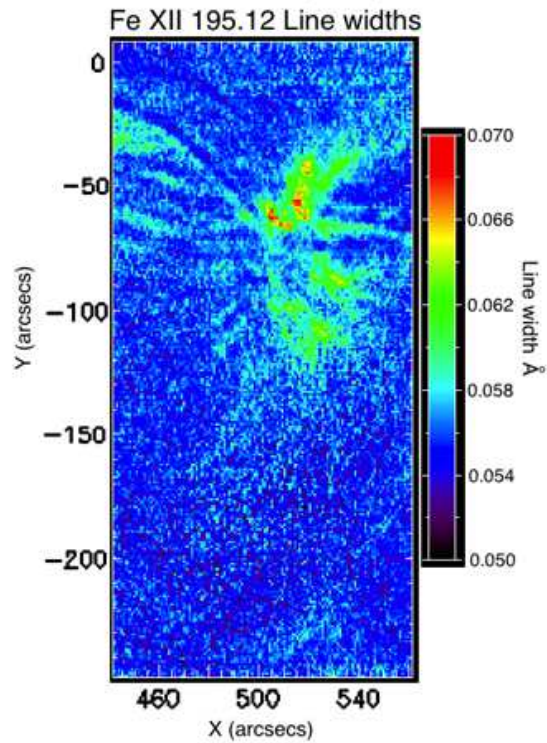}
\caption{{\it Hinode}/EIS  Fe {\small XII} 195.12 \AA\ image of spectral line broadening for the entire AR.}\label{fig:NT_195_full}
\end{center}
\end{figure}

\begin{figure}[h]
\begin{center}
\includegraphics[totalheight=4in]  {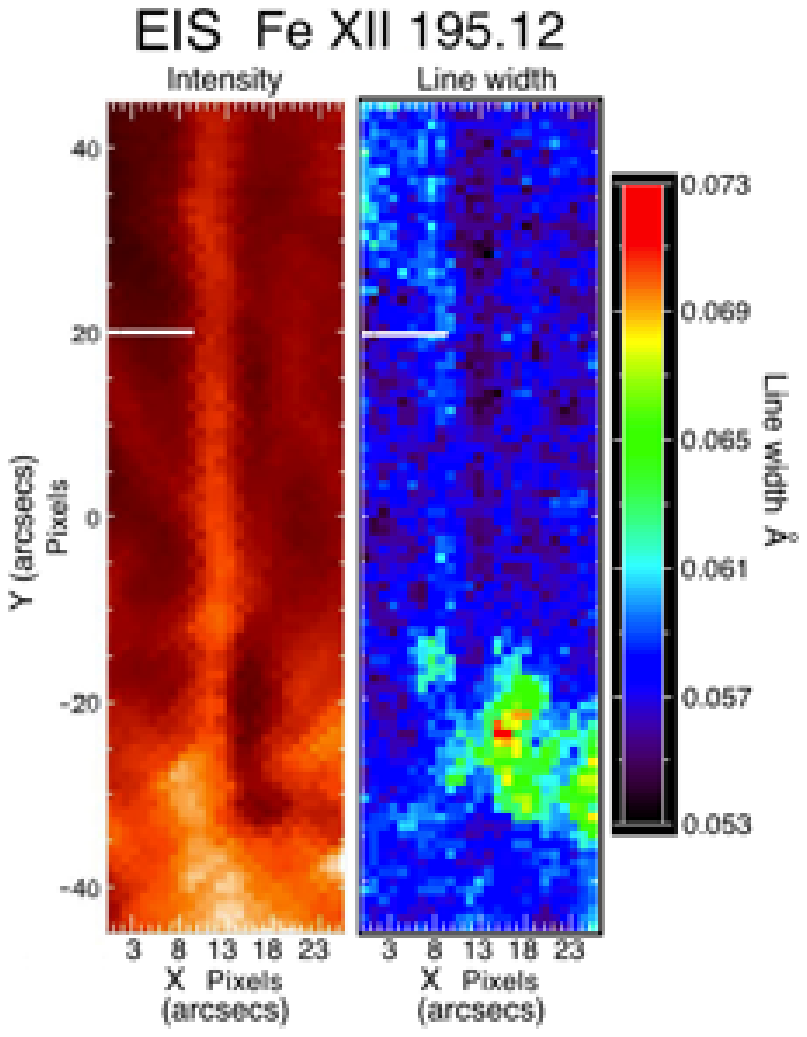}
\caption{{\it Hinode}/EIS  Fe {\small XII} 195.12 \AA\ images of the intensity (left) and spectral line broadening (right) for the isolated coronal loop structure. Where the loop is found in intensity, the loop structure is traced out by a region of minimal/low line broadening. } \label{fig:NT_195_loop}
\end{center}
\end{figure}

Figure \ref{fig:NT_195_full} shows a map of the Fe {\small XII} 195.12 line widths. The loop structure appears as a dark region of relatively small line width. Other loop structures that appear in the intensity image can be traced as extended regions of small line width.  Figure \ref{fig:NT_195_loop} is the line width map for the isolated loop structure.  The region of small line width is co-spatial with the dynamic loop seen in the intensity images. The area of greatest line widths lays directly adjacent to the foot point of the isolated loop structure.

\begin{figure}[h]
\begin{center}
\includegraphics[totalheight=5in]  {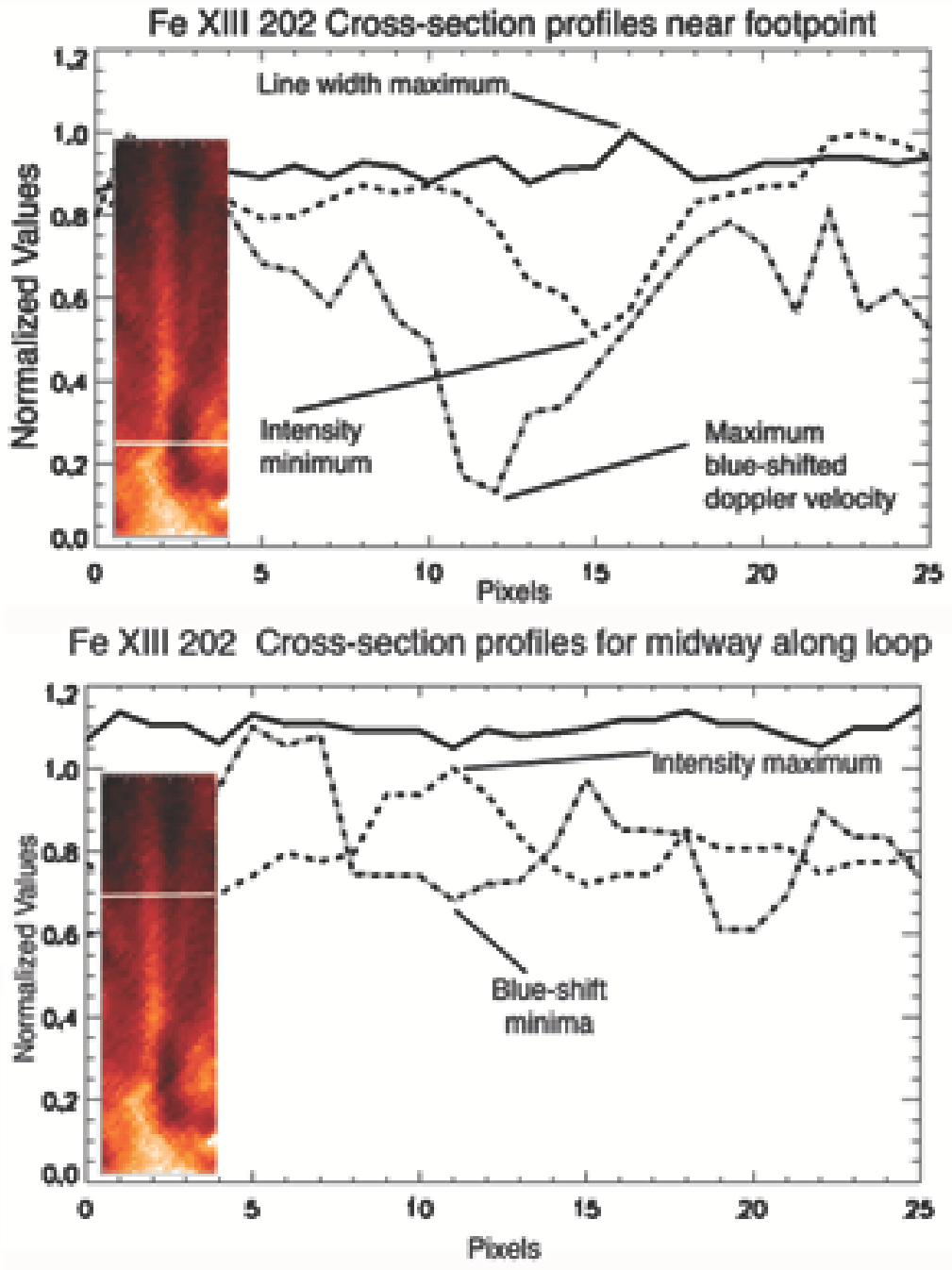}
\caption{{\it Hinode}/EIS  Fe {\small XII} 195.12 \AA\ intensity, line width, and Doppler velocity profiles for a cross-section near the foot point of the loop structure and for a region midway along the structure.  Solid line = line widths, dashed line = intensity, dash-dot = Doppler velocity. (The white line in the inset image indicates the cross-sectional cut used) The velocity range is -8 to 8 km sec$^{-1}$ for the footpoint region and -2 to 6 km sec$^{-1}$  for the section midway along the loop structure. The intensity, line width, and Doppler velocity values have been normalized and  in the lower plot the line width profile has been shifted +1.5 units in order to view the profile more clearly.}\label{fig:Cross_Profiles}
\end{center}
\end{figure}

\begin{figure}[h]
\begin{center}
\includegraphics[totalheight=3.5in]  {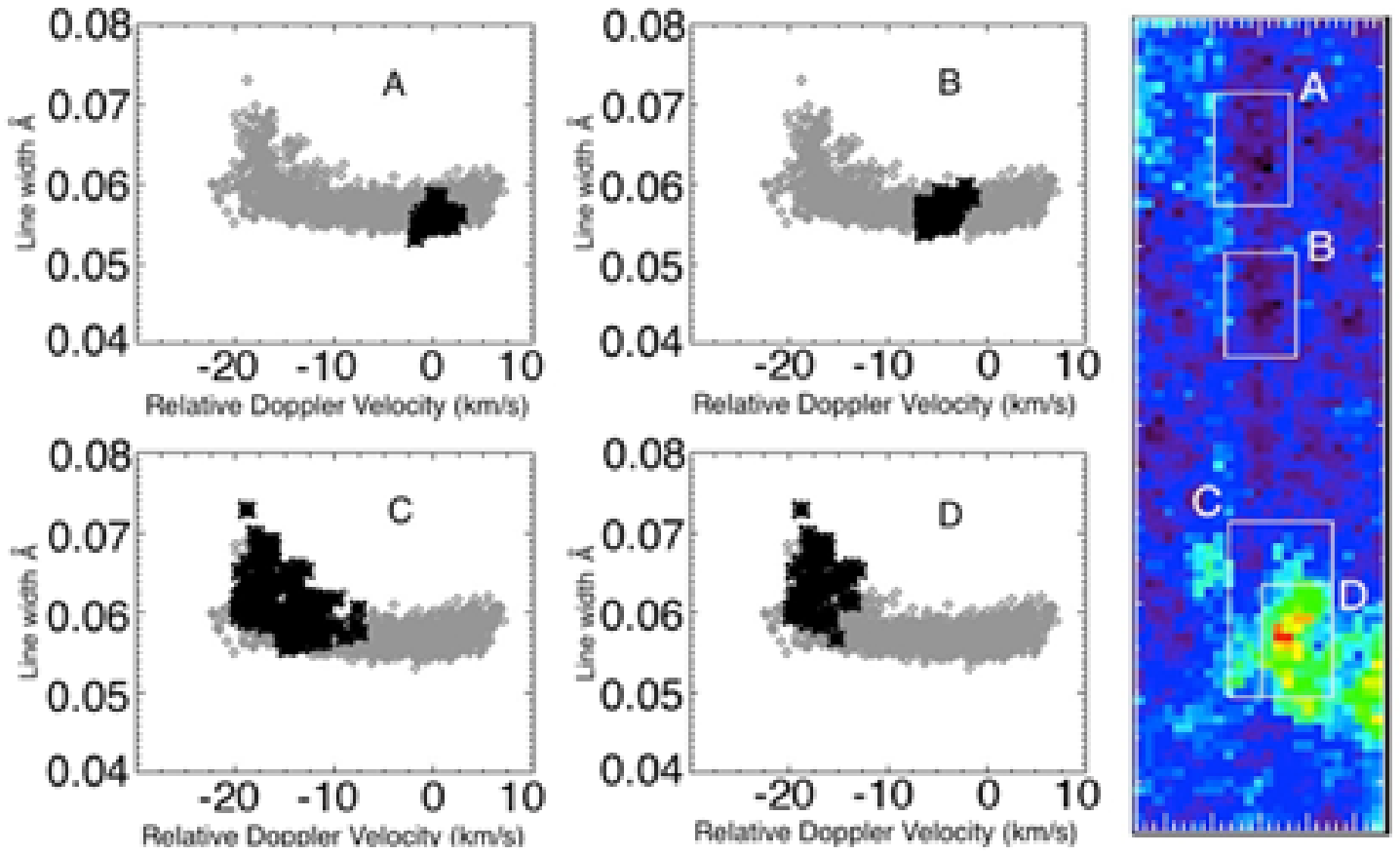}
\caption{{\it Hinode}/EIS  Fe {\small XII} 195.12 \AA\ line width map of the loop structure. The white boxes indicate the regions used for the scatter plots. The corresponding scatter plots on the left of the figure show the spectral line broadening. The grey diamonds show the scatter plot for the entire image. The black stars indicate data from the selected region.  Box A is near the apex of the loop structure. Box C includes the footpoint region and the nearby line broadening. Box D only includes the  line breoadening near the footpoint. }\label{fig:Scatterplots} 
\end{center}
\end{figure}

Figure  \ref{fig:Cross_Profiles} shows plots of line width profile, intensity and Doppler velocity for cross-sections of the isolated loop, near the foot point and midway along the loop structure. Figure \ref{fig:Cross_Profiles} indicates that the location of the increased line broadening is not co-spatial with the intensity or the blue shifted material in the loop structure. The maximum line broadening is seen where there is a minimum in intensity, while the minimum of the velocity profile seems to extend over both the intenisty and line width maxima. It appears as though the intensity and the line broadening are anti-correlated. In addition, there is no correlation of line broadening with the Doppler velocities within the structure. This can be seen in the scatter plots shown in Figure \ref{fig:Scatterplots}. The scatter plot for region A shows the material near the apex of the loop structure is not spectrally broadened nor is there a dominance of a Doppler shift in one direction. The material in region B is blue-shifted with very little spectral line broadening. Region C includes the footpoint region of the loop structure as well as the adjacent spectral line broadening. Scatter plot for region C shows a correlation of the Doppler velocities with the spectral line broadening. We can see that there are highly blue-shifed, highly broadened spectral lines for region C. When region C is decreased to include just the adjacent line broadening and very little of the nearby loop structure, we see that only the very blue-shifted and the very broadened data points remain. This demonstrates that no or very little spectral line broadening is occurring within the structure. The location of the maximum spectral line broadening is directly adjacent to the footpoint of the isolated loop structure.

 This analysis was performed with two other ion lines, Fe {\small XII}  186.854+186.887,  and Fe {\small XIII} 202.04 \AA~. The locations of the line broadenings are co-spatial in all three Fe {\small XII} ions 186.88, 195.12, and Fe {\small XIII} 202.04 \AA~. Fe {\small XII} 186.854+186.887 \AA~ showed consistently larger widths ( $\approx$ 0.01 \AA~) than Fe {\small XII} 195.12 and Fe {\small XIII} 202.04 \AA~. This is most likely be due to the blending of the two Fe {\small XII} lines 186.854 and 186.887 \AA~. The line width measurements of the unblended Fe {\small XIII} 202.04 \AA~ and the (corrected for blending) Fe {\small XII} 195.12 agree. The Fe {\small XIII} 202.04 \AA~ line shows the same anti-correlation with intensity in the loop structure and maximum line broadening adjacent to footpoint the loop structure, but the maximum line widths of Fe {\small XIII} are found to be  $\approx$ 0.02 \AA~ smaller than those of the corrected Fe {\small XII} 195.12 \AA~.

\subsection{Loop Cooling Time}
The determination of physical parameters associated with the loop structure allows for comparison to estimates of the conductive and radiative cooling times of a loop derived from the energy equations (Rosner, Tucker, and Vaiana 1978). 

Estimates of the radiative and conductive cooling times (see Section 4.2 of Scott, Martens, and Cirtain, 2008) were found to be 16 minutes and  8 hours respectively. The estimated time including both mechanisms can crudely found by evaluating the inverse of the addition of the inverted times  (similar to adding resistances in parallel)  which is found to be 968 seconds ($\approx\ 16$ min).  It is important to realize that these cooling time estimates are \emph{characteristic} cooling times, found by independently evaluating the radiative and conductive terms in the energy equation at a single temperature (Cargill, Mariska, and Antiochos, 1995).
 
An estimate of the total cooling time based on integration of both the radiative and conductive terms along with the temperature response of the instrument was calculated. A method similar to that put forth by \v{S}vestka (1987) was used to evaluate the cooling of the plasma through the instrument filters. The intitial temperature log(T) = 6.5 and final temperature log(T) = 5.3 were found from the STEREO and TRACE 171 \AA~ filter temperature response functions. The time to cool between these temperatures, with this method is 41 minutes.

\subsection{Summary of Results}

A coronal structure appears in a range of temperatures log(T) = 6.0-6.3 K. This structure was isolated from the background and investigated for the lifetime of the structure (4-5 hours). The structure was observed at the resolution limits of the instruments and therefore was not adequately resolved. However estimates of the physical structure of the loop were made using EIS data. The loop appeared to decrease in density $10^{10}$ to $10^{9.3}$ cm$^{-3}$ from the footpoints to the apex. The halflength of the loop calculated from the filter images was found to be 69 Mm with the upperbound width ranging from 1-1.5 Mm. Filling factors indicate the emitting structure remains unresolved and may have an expansion from 50 to 100 km along its length. The loop appears for longer than the calculated cooling times and is overdense by an order of magnitude with respect to RTV scaling laws (Rosner, Tucker, Vaiana, 1978). Lightcurves show varying brightness in 171 and 195 \AA~ over the entire lifetime with varying durations. The intensity curves and calculated cooling times show that the loop is not simply cooling through the filter passbands. The LOS relative Doppler images indicate flows along the loop structure during the time EIS observed the region. The blueshifts from high temperature (T $> 1.25$) ions are present in the outer areas of the AR where EIS data indicates there are low density, low intensity regions and  redshifted material from lower temperature ions outlines the footpoint regions of cooler structures. These results, in general,  agree with the observations of Del Zanna (2008), but the presence of a distingushable blueshifted loop structure that is brighter and more dense than the background suggests that upflows may be associated with bright coronal features as well. The spectral line broadening was shown to be anti-correlated with the intensity in the loop structure. The maximum spectral line broadening was found very near the footpoint of the investigated loop structure.

\section{CONCLUSIONS}

A non-flaring active region loop structure is observed during its lifetime with a combination of Hinode/EIS, TRACE, and STEREO. The focus has been to make the key measurements of the parameters that are needed to constrain and allow comparison to different heating models. The lifetime, temperature structure, length, width, density, spectral line broadening,  and LOS relative Doppler shifts have been measured. The long lifetimes and varied duration brightenings are difficult to reconcile with any loop heating/cooling model. The lightcurves do not show that the structure is simply cooling through the filter passbands. If considering only the visible lifetime of the structure it could appear that there is a kind of static equilibrium, since the structure can be seen for multiple hours. However the loop is overdense with respect to static scaling laws and shows dynamic flux variations, on the 10-20 min. and 1-4 hour time scales, during the visible lifetime of the structure. The investigation of an isolated structure shows loop structures that are bright only within a range of temperatures log(T) = 6.0-6.3. These loops vary in intensity and exist for longer than estimated cooling times. This suggests a quasi-steady heating mechanism that is able to produce both short and long duration brightenings. Quasi-steady in the respect that something is energizing the structure enough to maintain its appearance for longer than the calculated cooling times for a single loop. It may be that the mechanisim energizes multiple sub-resolution loops to 1-2 MK temperatures and maintain an apparent structure. The sub-resolution loops then cool so rapidly that the loops do not become visible above the background in the lower temperatures. 

Relating the flows in different temperature regimes becomes even more difficult. While the general flow structure agrees with previous observations, relative blue shifts only delineate the particular structure analyzed. There are structures that seem to share the same footpoint region in the higher temperature images, yet these higher temperatures do not delineate any particular coronal structure. This points to the blueshift being associated solely with the analyzed loop, linking the brightening with a relative blueshift. This may suggest a different evolution for loops with these properties.

The structure's apparent width remains constant along the length and throughout the lifetime. These measurements lie close to the resolution limits of the observing instruments so they cannot be considered completely elemental strands. Filling factors of 0.047 and  0.09 along the structure support the idea that the emitting structure has not been resolved.

Hot steady plasma loops are generally associated with the cores of active regions, while the periphery is associated with cooler longer loops. We observe, in this active region, hot fully illuminated loop structures in the core with the footpoints of longer loops in the periphery seen in (a wider range of temperatures) cool and warm temperatures. These cooler loops display redshifted material that is expected from coronal rain. We find a ``warm'' loop in the core region of the active region that is evolving. We have shown that there are relative differences in the flows of this short length structure with a narrow temperature structure and that of the longer length structure with a broader temperature structure. We observe blueshifted material that delineates an isolated coronal loop structure. This material is shown to be dynamic, relatively dense and bright against the background, which suggests that bright, dense material need not always be associated with redshifts. This demonstrates that dynamically evolving coronal structures can be composed of blueshifted material. 

The absence of spectral line broadening in the loop structure is an interesting result. This study confirms a suggestion of Doschek {\it et al.} (2007) that the line widths may be larger in regions adjacent to well-defined loops. We have found that our isolated coronal loop structure does not show increased line broadening along its length. In addition, we find the maximum line widths near the footpoint of the loop structure. The loop structure exhibits blue-shifted wavelength centroids. That could indicate, as Doschek {\it et al.} (2007) pointed out, that the loop structures may be embedded in a descending cloud of plasma with large nonthermal velocities. This scenario is appealing since it may present a clue to the observations that high temperature images appear ``fuzzy'' relative to lower temperature images. The descending cloud scenario could suggest that the images are simply ``cloudy'' relative to lower temperature images. 

 A descending cloud of plasma does not explain the observation that the maximum line broadenings are found directly near the footpoint region of a loop structure. The location of the maximum line widths indicates that the phenomena creating the line broadening may be concentrated at low altitudes near loop footpoints. The absence of spectral line broadening in the loop structure may be due to the structure masking or shielding the increase in line width or it may be that the higher  density structure is reducing the line broadening. This would imply that the loop structure is between the observing mechanism and the source of the line broadening.

A coronal loop structure has been isolated and investigated. This loop structure dynamically brightens with two different time scales and is overdense according to current loop models. The loop structure does not show the general characteristics of low intensity, low density, high blue shifted, and large spectral line widths associated with outflow regions of active region. Instead the loop structure is found to have a large intensity that is co-spatial with blue-shifted, line width reduced spectral profiles. The line width maxima are found adjacent to the footpoints of the loop structure. These observations point to multiple processes at work brightening and heating the corona. They also suggest that the bulk of the line broadening occurs at low altitudes directly adjacent to loop structure footpoints. 

In this paper we analyze in great detail a single loop, rather than collect data on an ensemble of loops. The reason for this focus is that we have data of unsurpassed quality for this loop: a unique sequence of 12 hours of observations during which the loop retains its identity for 5 hours, a very clear and complete loop outline in the TRACE images, a persistent flow within the loop, and excellent spectra that clearly distinguish the loop from its surroundings in terms of intensity, Doppler velocity, density, and spectral line-broadening. As a result of this we have been able to characterize the 3D loop geometry, spectral line broadening, temperature, density, and flow velocity inside loop for an extended period of time with great accuracy.  Having these physical measurements in hand is of great use in the verification of theories for coronal heating.

\acknowledgements
We would like to thank the anonymous referee for his/her diligence and suggestions that improved this article.
{\it Hinode} is a Japanese mission developed and launched by ISAS/JAXA, collaborating
with NAOJ as a domestic partner, and NASA (USA) and STFC (UK) as international partners.
It is operated by these agencies in co-operation with ESA and NSC (Norway).

\end{article} 
\end{document}